\documentclass[aps,prd,twocolumn,groupedaddress, showpacs]{revtex4} 

\usepackage{graphicx}

\begin{document}

\title{Constraints on amplitudes of curvature perturbations from \\ primordial black holes}

\author{Edgar Bugaev}
\email[e-mail: ]{bugaev@pcbai10.inr.ruhep.ru}

\author{Peter Klimai}
\email[e-mail: ]{pklimai@gmail.com}
\affiliation{Institute for Nuclear Research, Russian Academy of
Sciences, 60th October Anniversary Prospect 7a, 117312 Moscow,
Russia}




\begin{abstract}

We calculate the primordial black hole (PBH) mass spectrum
produced from a collapse of the primordial density fluctuations in the early
Universe using, as an input, several theoretical models giving the
curvature perturbation power spectra ${\cal P}_{\cal R}(k)$ with
large ($\sim 10^{-2} \div 10^{-1}$) values at some scale of
comoving wave numbers $k$. In the calculation we take into account the explicit dependence
of gravitational (Bardeen) potential on time.
Using the PBH mass spectra, we further calculate the
neutrino and photon energy spectra in extragalactic space from
evaporation of light PBHs, and the energy density fraction contained in PBHs today (for heavier PBHs).
We obtain the constraints on the model parameters using available experimental data
(including data on neutrino and photon cosmic backgrounds).
We briefly discuss the possibility that the observed 511 keV line from the Galactic center is produced
by annihilation of positrons evaporated by PBHs.

\end{abstract}

\pacs{98.80.-k, 04.70.-s} 

\maketitle

\section{Introduction}

It is well known that for a sufficient production of primordial
black holes (PBHs) in the early Universe
\cite{Zeldovich1967, Hawking1971, Carr:2005zd, Khlopov:2008qy}
the spectrum of the density
perturbations set down by inflation must be ``blue'', i.e., it
must have more power on small scales. This implies that the
spectral index of the scalar perturbations must be larger than 1,
in strong contradiction with the latest WMAP results
\cite{Spergel:2006hy, Dunkley:2008ie, Komatsu:2008hk}. In
particular, standard inflationary models of hybrid type, in which the
inflaton is trapped in a local minimum of the potential and which
predict blue spectra seem to be excluded as a possible source of
PBHs.

Such a conclusion is correct, however, only in rather special case: namely, it is
based on the prediction of slow-roll single field inflationary scenario, according to which
the power spectrum of curvature perturbations is nearly scale-invariant, i.e.,
the spectral index $n$ is close to unity and the variation in the spectral index
$dn/d \log k$ is small.

Although a prediction of the approximate scale invariance of the primordial power
spectrum is a necessary requirement to any inflationary model, some deviations from pure scale
invariance are consistent with the observational data. These deviations are described
by adding localized features to the primordial spectrum (see, e.g., \cite{Hoi:2007sf} and references
therein) and/or by introducing spectral features modifying a single power law. Models
with such peculiarities (sometimes called broken-scale-invariant (BSI) models) were
proposed, in main aspects, in eighties \cite{Starobinsky:1986fxa, Kofman:1985aw,
Kofman:1986wm, Silk:1986vc, Kofman:1988xg, Salopek:1988qh}. Such models generally
include, in addition to the usual inflaton field, other scalar fields driving
successive stages of inflation and triggering phase transitions.

Evidently, the BSI models of inflation could predict, generically, the essential production
of primordial black holes at small and medium scales. In particular, in \cite{Starobinsky:1992ts}
the inflationary model with a steplike power spectrum based on a nonanalyticity in
the inflaton potential was proposed and in subsequent works the inflationary potentials of
such kind were used for making predictions of PBH production \cite{Ivanov:1994pa,
Ivanov:1997ia, Bullock:1996at}.

A step feature in the potential is an effective field theory description of a phase transition,
so, in a more realistic approach an inflationary model should certainly involve more than one
scalar field, and such rapid phenomena as phase transitions \cite{Starobinsky:1998mj}. Second order
phase transitions during inflationary expansion had been first considered in
\cite{Kofman:1986wm, Kofman:1988xg} in models with two scalar fields. In scenarios of
such type, during a short stage, corresponding to the beginning of a phase transition,
the mass of the trigger field becomes negative, and adiabatic perturbations are exponentially
amplified resulting in the formation of a narrow spike in the primordial spectrum, and, as
a consequence, in a copious production of PBHs \cite{GarciaBellido:1996qt, Randall:1995dj}.
There are many multiple field scenarios predicting the existence of spike- or bumplike
features in the primordial spectrum (e.g., supersymmetric double hybrid models \cite{Lesgourgues:1999uc},
multiple inflation models based on supergravity \cite{Adams:1997de}, etc). Some of these models
are specially constructed to predict efficient PBH production \cite{Yamaguchi, Kawasaki}.

Further, large curvature perturbations (especially on subhorizon scales) leading to features in the primordial
spectrum and possible PBH production, can arise in multiple field scenarios at the end of
inflation (during preheating era) or between two consecutive stages of inflation, as a result of
parametric resonance or tachyonic instability. Estimates of PBH production possibilities had been
obtained in  \cite{Green:2000he, Bassett:2000ha, Finelli:2001db} for two-field models
and in \cite{Suyama:2006sr} for models with a self-interacting scalar field. The more
complicated model (containing three fields), based on supergravity, is considered in
\cite{Kawasaki}.

An existence of the narrow spikes in the primordial spectrum is possible not only in multiple
field inflationary scenarios. Such a feature can, in principle, exist even in single field
models (see, e.g., \cite{Saito:2008em, Bugaev:2008bi}). If, in particular, the inflationary potential
has an unstable maximum at origin (e.g., the double-well potential) then, with some
fine-tuning of parameters and initial conditions, the inflation process may have two stages,
with a temporary stay at the maximum, that may lead to the corresponding peak in the
primordial spectrum and, depending on the amplitude of the peak, to the PBH production.


Another example of an inflationary model predicting large amplitudes of the density perturbations
at small scales is the inflationary model with the running mass potential. More generally,
large amplitudes of the primordial curvature perturbation spectrum at small scales are
possible in models of "hilltop hybrid inflation" \cite{Kohri:2007gq}. Potentials of these
models have concave-downward form at cosmological scales (corresponding to red primordial
spectrum as required by observations), but can be much flatter near the end of inflation.
Such a modification of a standard hybrid inflation scenario is discussed, e.g., in
\cite{NeferSenoguz:2008nn, Rehman:2009wv}. The running mass potential belongs to the class
of hilltop potentials; in models with such potentials, due to spectral index scale-dependence,
the amplitude of the perturbation spectrum at small scales depends on the value of $dn/ d\log k$
at cosmological scales \cite{Kohri:2007gq}. It is shown in our previous work
\cite{Bugaev:2008bi} that possibilities of noticeable production of PBHs in running mass
model are still open in spite of the rather severe experimental constraints
\cite{Kohri:2007qn, Peiris:2008be} on the value of the spectral index running.

In this paper we limit ourselves to the situations where PBHs form from the
density perturbations, induced by quantum vacuum fluctuations during inflationary expansion.
The details of the PBH formation had been studied in \cite{Carr:1975qj, Khlopov:1980mg}, the
astrophysical and cosmological constraints on the PBH density had been obtained in many
subsequent works (see, e.g., the recent review \cite{Khlopov:2008qy}). The order of
magnitude of the corresponding constraint on the value of the density perturbation amplitude
(for the PBH mass region $10^{11} - 10^{18}$ g, which we are interested in) is well known
\cite{Carr:1994ar}, but, if the primordial spectrum contains the peaklike feature, the concrete
value of the PBH constraint clearly depends on the parameters characterizing the form of
this feature (in particular, on the width of the peak). Such an information may be rather
useful for the model makers.

One must note that the formation of PBHs in models with the primordial spectrum having the
features had been studied earlier, in works \cite{Yokoyama:1998xd, Blais:2002gw}. In
\cite{Yokoyama:1998xd} the case of the spectrum that is sharply peaked on a single mass
scale was considered. The $\beta(M)$ function (the probability of a region of mass $M$ to
form a PBH) was calculated and it was shown that the form of this function depends on
the type of the gravitational collapse. In \cite{Blais:2002gw} two cases were studied: a pure
step in the primordial spectrum and the spectrum produced in an inflationary model \cite{Starobinsky:1992ts}
with a jump in a first derivative of the inflaton potential at some scale. It was shown that the
corresponding $\beta$-functions have the pronounced bumps and, in connection with
these bumps, authors of \cite{Blais:2002gw} discuss the possibility that PBHs can be
a significant part of dark matter.

In the present work we reconsider the problem of constraining the
power spectrum of the primordial fluctuations (with accent on the peaklike features)  calculating the process
of the formation of PBHs having small masses ($\sim 10^{11} -
10^{18}$g). Products of evaporation of these PBHs contribute, in
particular, to extragalactic photon and neutrino diffuse
backgrounds (which are measured experimentally), and the constraints are calculated using the
standard procedure \cite{Page:1976wx}.
We do not calculate the $\beta$-functions preferring
to constrain the power spectrum directly. The concrete constraints are obtained for two special
cases (the power spectrum with a peak and the spectrum of the running mass model), although
the general formulas of Sec. \ref{sec-form} and \ref{sec-nu-gamma} can be used for any form of the
spectrum.

At the end of the paper we use our approach for a checking, once more, the idea proposed in
\cite{Frampton:2005fk, Bambi:2008kx}, namely, we study the possibility that evaporating PBHs
(having mass spectrum calculated with formulas of Sec. \ref{sec-form})
cluster in the Galactic center and produce the observed $511$ keV photon line.

The plan of the paper is as follows. In Sec. \ref{sec-form} we present the general formalism of the PBH mass
spectrum calculation with taking into account the explicit time dependence of the gravitational potential.
The results of such a calculation for the concrete example, in which the curvature perturbation spectrum has
a sharp peak, are given.
In Sec. \ref{sec-nu-gamma} we give some details of the calculation of the  neutrino and the photon
extragalactic diffuse spectra from the PBH evaporations. In Sec.
\ref{sec-constr-peak} and Sec. \ref{RM-sec} the constraints
on the curvature spectrum from PBHs are given, for the model of ${\cal P}_{\cal R}(k)$ with a peak
and for the running mass model, respectively.
In Sec. \ref{sec-DM} we consider the possibility of the explaining the observed $511$ keV photon line
from the Galactic center by PBHs clusterizing there.  Section \ref{sec-concl} contains our conclusions.

\section{PBH mass spectrum calculation}
\label{sec-form}

\subsection{General formula}

The calculation of PBH mass spectrum in Press-Schechter formalism
\cite{PS} is based on the expressions \cite{Kim:1996hr, Kim:1999xg, BugaevD65}
\begin{eqnarray}
n_{BH}(M_{BH}) d M_{BH} = \;\;\;\;\;\;\;\;\;\;\;\;\;\;\;\;\;\;\;\;\;\;\;\;\;\;\;\;\;\;\;\;\;\;\;\;\;\;\;\;\;\;\;\;\;
 \label{nBHMBH}\\
= \left \{ \int n(M, \delta_R) \frac{d\delta_R}{d\delta_R^H} \frac{dM}{dM_{BH}}
d\delta_R^H \right \}dM_{BH} \nonumber \;,
\\
\label{nBHMBH2}
n(M, \delta_R) = \sqrt{\frac{2}{\pi}} \frac{\rho_i}{M}
\frac{1}{\sigma_R^2} \left| \frac{\partial \sigma_R}{\partial M}
\left( \frac{\delta_R^2}{\sigma_R^2} -1 \right) \right|
e^{-\frac{\delta_R^2}{2\sigma_R^2}}.
\end{eqnarray}

Here, the following notations are used: $\delta_R$ is the initial
density contrast smoothed on the comoving scale $R$, $M$ is the smoothing
mass, $\sigma_R(M)$ is the mean square deviation (the mass variance),
\begin{equation}
\sigma_R^2(M) = \int \limits_{0}^{\infty} {\cal P}_\delta(k) W^2(kR)
\frac{dk}{k}, \label{SigR}
\end{equation}
${\cal P}_\delta(k)$ is the power spectrum of primordial density
perturbations, $W(kR)$ is the Fourier transform of the window function (in this work we use
the gaussian one, $W(kR) = \exp (- k^2 R^2/2 )$ ),
$\rho_i$ is the initial energy density.
It is assumed that the process of reheating is very short in time, so, the end
of inflation practically coincides with a start (at $t=t_i$) of the radiation era.

Fourier transform of the (comoving) density contrast is
\begin{equation}
\delta_k(t) = -\frac{2}{3} \left(\frac{k}{aH} \right)^2 \Psi_k(t),
\label{dkt}
\end{equation}
where $\Psi_k$ is the Fourier transform of the Bardeen potential.
In the approximation $\Psi_k = {\rm const}$, we have, for the radiation dominated epoch, $\delta_k \sim a^2$,
that is the well-known result \cite{Kolb}. In this work, however, we will not use this approximation
and, in opposite, we will explicitly take into account the time dependence
of the Bardeen potential.

The power spectrum of the density perturbations, calculated at some moment of time, is
\begin{equation} \label{Pdeltak}
{\cal P}_\delta(k, t) = \left[ \frac{2}{3} (k \tau)^2  \right]^2 {\cal P}_{\Psi}(k, t),
\end{equation}
where $\tau$ is the conformal time ($\tau=(aH)^{-1}$ for the radiation
epoch).

The comoving smoothing scale, $R\equiv
1/k_{R}$, is connected with the smoothing mass $M$ by the expression
\begin{equation}
\label{kflaiHi} %
\Big( \frac{M}{M_i} \Big) ^{-2/3} = \frac{k_{R}^2}{(a_i H_i)^2},
\end{equation}
where $M_i$, $a_i$ and $H_i$ are the horizon mass, cosmic scale factor and Hubble parameter at the
moment $t_i$. Eq. (\ref{kflaiHi}) immediately comes from two simple formulas: horizon
mass at $t_i$ is given by
\begin{equation}
M_i = \frac{4\pi}{3}  t_i ^3 \rho_i,
\label{Mirhoi}
\end{equation}
and the smoothing mass is
\begin{equation}
M = \frac{4\pi}{3} \frac{a_i^3}{k_R^3} \rho_i.
\end{equation}
From these relations, one obtains
\begin{equation}
\frac{M}{M_i} = k_R^{-3} \left( \frac{a_i}{t_i} \right)^3 ,
\end{equation}
and putting $H_i=1/t_i$ we come to Eq. (\ref{kflaiHi}).

For the practical use of formulas (\ref{nBHMBH}, \ref{nBHMBH2}) one must express the PBH
mass $M_{BH}$ through the smoothing mass $M$ and the density
contrast at the moment of the collapse (which is approximately equal to the density
contrast at horizon crossing, $\delta_R^H$).

It follows from Eq. (\ref{dkt}) that the smoothed density contrast at the initial moment of time,
$\delta_R$, and at the time of horizon crossing, $\delta_R^H$,  are connected by the expression
\begin{equation}
\frac{\delta_R}{\delta_R^H} \approx
\frac{\delta_{k^*}(t_i)}{\delta_{k^*}(t_h)} = \left(
\frac{\tau_i}{\tau_h} \right)^2
\frac{\Psi_{k^*}(\tau_i)}{\Psi_{k^*}(\tau_h)} \; .
\end{equation}
Here we use the approximation according to which the smoothed density contrast is proportional
to its Fourier transform at some characteristic value of $k$, $k=k^*$. If the power spectrum
${\cal P}_\delta(k)$ increases monotonically with $k$, then, evidently,  $k^* \approx k_R = 1/R$.
If the spectrum has a maximum at some value of $k$, $k=k_0$, the reasonable estimate for $k^*$ is
\begin{eqnarray}
k^* \approx %
\left\{  \begin{array} {l}
  k_R \;\; , \;\; {\rm for } \;\; k_R < k_0 \; ,\\
  k_0 \;\; , \;\; {\rm for } \;\; k_R > k_0 \; .
\end{array} \right.
\end{eqnarray}

\subsection{Gravitational collapse models}

The connection between values of the smoothing mass $M$, density contrast $\delta_R^H$
and PBH mass $M_{BH}$ can be expressed in the general form
\begin{equation}
M_{BH} = f(M, \delta_R^H; M_i) .
\end{equation}
The concrete expression for the function $f$ depends on the
model of the gravitational collapse. In the model of the standard
spherically-symmetric collapse the connection is quite simple:
\begin{equation}
\label{MBHsph} %
M_{BH} =  (\delta_R^H)^{1/2} M_h.
\end{equation}
Here, $M_h$ is the horizon mass at the moment of time, $t=t_h$,
when regions of the comoving size $R$ and smoothing mass $M$ cross horizon. According to Carr
and Hawking \cite{CarrHawking}, $1/3 \le \delta_R^H \le 1$. The derivation
of Eq. (\ref{MBHsph}) is given in the Appendix \ref{app1}.

The horizon mass at the moment when a region of the comoving size $R$ crosses horizon is
\begin{equation}
M_h = \frac{4 \pi}{3} (aR)^3 \rho(a)
\end{equation}
(at this moment, $a/k_R=H^{-1}$). Using (\ref{Mirhoi}), one obtains
\begin{equation}
M_h = M_i  H_i \frac{a}{k_R} = M_i \frac {(a_i H_i)^2} {k_R^2}.
\end{equation}
Finally, using (\ref{kflaiHi}), the connection between $M_h$ and $M$ is derived \cite{BugaevD65}:
\begin{equation}
\label{MhMiM} %
M_h = M_i^{1/3} M^{2/3} .
\end{equation}
This equation connects horizon mass at the moment of time $t_h$ when the perturbed
region crosses horizon with the initial (at $t_i$) horizon mass and the initial (at $t_i$)
comoving mass $M$. It is seen that the comoving mass decreases with time ($M_h<M$)
(see, e.g., \cite{Green:1997sz}).

From (\ref{MBHsph}, \ref{MhMiM}) one has the expression for the
function $f$ for the Carr-Hawking collapse:
\begin{equation}
\label{fCH} %
f(M, \delta_R^H; M_i) = (\delta_R^H)^{1/2} M^{2/3} M_i^{1/3}.
\end{equation}

In the picture of the critical collapse \cite{NJ} the
corresponding function is
\begin{equation}
\label{fcrit} %
f(M, \delta_R^H; M_i) = k_c (\delta_R^H-\delta_c)^{\gamma_c}
M^{2/3} M_i^{1/3},
\end{equation}
where $\delta_c$, $\gamma_c$ and $k_c$ are model parameters. In
this work we will accept the following set of parameters, which is
in agreement with the recent calculations \cite{Musco, Musco:2008hv}:
\begin{equation}
\label{parcrit} \delta_c=0.45, \;\;\;\; \gamma_c=0.36, \;\;\;\;
k_c=4.
\end{equation}

One should note that the neglect of a time dependence of the gravitational
potential greatly simplifies a calculation of the PBH mass spectrum because
the smoothed density contrast $\delta_R$ enters the PBH mass spectrum
expression only through the ratio $\delta_R/\sigma_R$. In the limit of
the constant $\Psi_k$ one has, from Eqs. (\ref{MhMiM}) and (\ref{kflaiHi}),
\begin{equation}
\delta_R \approx \frac{M_i}{M_h} \delta_R^H = \frac{k_R^2}{a_i^2 H_i^2} \delta_R^H .
\label{e1}
\end{equation}
In the same limit, an expression for the mass variance is
\begin{eqnarray}
\sigma_R^2 \approx \left( \frac{k_R}{a_i H_i} \right)^4 \int \frac{4}{9} (kR)^4 {\cal P}_\Psi(k)
W^2(kR) \frac{dk}{k} \equiv  \nonumber \\
\equiv \left( \frac{k_R}{a_i H_i} \right)^4 \sigma_H^2 .\;\;\;\;\;
\label{e2}
\end{eqnarray}
It follows from Eqs. (\ref{e1}) and (\ref{e2}) that
\begin{equation}
\frac {\delta_R^2} {\sigma_R^2} \approx \frac {(\delta_R^H)^2} {\sigma_H^2}.
\end{equation}
This ratio is constant with time. It is clear from these arguments that a time dependence
of $\Psi_k$ (given, e.g., in Fig. \ref{psi-fig}) will affect the form of the PBH mass
spectrum (see Sec. \ref{RM-sec}).

\subsection{Time dependence of the gravitational potential}

For a calculation of a time dependence of the gravitational potential we use the
approach suggested in works of Ref. \cite{Lyth:2005ze}. We assume that reheating is rapid and,
correspondingly, the moment of the end of inflation coincides with the moment of a beginning
of the radiation epoch. The only difference from the consideration of \cite{Lyth:2005ze} is
that we do not assume, in general, the validity of the slow-roll approximation at the period near the end
of inflation.

The general expression for the Fourier transform of the Bardeen potential at the radiation epoch is
\begin{equation}
\Psi_k^{\rm (rad)}(\tau) = A_k f_A(x) + B_k f_B(x)\; ,
\end{equation}

\begin{equation}
f_A(x) = \frac{1}{x^3} (x \cos x - \sin x) \; ,
\end{equation}

\begin{equation}
f_B(x) = \frac{1}{x^3} (x \sin x + \cos x) \; ,
\end{equation}
where $x=c_s k \tau$, $c_s^2 = \dot p/ \dot \rho = 1/3$ is the sound speed. The connection between
$\Psi_k$ and Fourier components of the curvature perturbation $\cal R$ in gauge invariant cosmological
perturbation theory is given by two expressions (see, e.g., the review \cite{Straumann:2005mz}):
\begin{widetext}
\begin{equation} \label{dotRk}
\dot {\cal R}_k(\tau) = \frac{2}{3} H \left(
\frac{k}{aH} \right) ^2 \frac{\Psi_k(\tau)}{1+w},
\end{equation}
\begin{eqnarray} \label{c2}
\frac{2}{3 H} \dot \Psi_k(\tau) + \frac{5+3 w}{3}
\Psi_k(\tau) = 
-(1+w) {\cal R}_k(\tau) \;.
\end{eqnarray}
For obtaining the coefficients $A_k$, $B_k$ we need two conditions for $\Psi_k(\tau)$.
The first one is the Eq. (\ref{dotRk}) calculated for the end of inflation, when
${\cal R}_k = {\cal R}_k^{\rm (inf)}$,
$\Psi_k = \Psi_k^{\rm (inf)}$,
$\tau=\tau_i$, and $aH=a_i H_i$. The second one is the Eq. (\ref{c2}) calculated for the beginning
of the radiation era when
${\cal R}_k = {\cal R}_k^{\rm (rad)}$,
$\Psi_k = \Psi_k^{\rm (rad)}$, and
$w=w_0=1/3$. The junction conditions at the transition time are
[\cite{Deruelle:1995kd}, \cite{Martin:1997zd}, \cite{Lyth:2005ze}]:
\begin{equation}
\Psi_k^{\rm (inf)}(\tau_i) = \Psi_k^{\rm (rad)}(\tau_i) \; , \label{c3}
\end{equation}
\begin{equation}
{\cal R}_k^{\rm (inf)}(\tau_i) = {\cal R}_k^{\rm (rad)}(\tau_i) \;. \label{c4}
\end{equation}

Using Eqs. (\ref{dotRk}, \ref{c2}) and (\ref{c3}, \ref{c4}), one obtains for the coefficients
$A_k$, $B_k$ the following two conditions:
\begin{eqnarray} \label{c5}
A_k f_A(x_i) + B_k f_B(x_i) = 
\frac{3}{2 H_i} \left(
\frac{k}{a_i H_i} \right) ^{-2} (1+w) \; \dot {\cal R}_k^{\rm
(inf)}(\tau_i) \; ,
\end{eqnarray}
\begin{eqnarray} \label{c6}
\frac{2 c_s k}{3 a_i H_i} \left[ A_k f_A'(x_i)  + B_k
f_B'(x_i) \right] + 
\frac{5 + 3 w_0}{3} \left[ A_k
f_A(x_i) + B_k f_B(x_i) \right] = 
-(1+w_0) {\cal R}_k^{\rm (inf)}(\tau_i) \; .
\end{eqnarray}

Two inputs in these expressions, ${\cal R}_k^{\rm (inf)}(\tau_i)$ and
$\dot {\cal R}_k^{\rm (inf)}(\tau_i)$, can be determined by a numerical calculation
(see, e.g., \cite{Bugaev:2008bi}) using the concrete inflation model.

In the present paper, we assume, for simplicity, that the right-hand side (rhs) of Eq. (\ref{c5}) is
equal to zero. It does not mean, however, that we exclude a consideration of non-slow-roll
models because the rhs of Eq. (\ref{c5}) contains, except of the term $(1+w)$, also the time
derivative of ${\cal R}(k)$ which is typically small due to freezing out of ${\cal R}(k)$ outside of
horizon  (if the comoving size $k^{-1}$ crosses horizon some time before the end of inflation,
see, e.g., \cite{Bugaev:2008bi}). The general case when violations of slow-roll are continuing up to the end
of inflation (and when, correspondingly, the rhs of Eq. (\ref{c5}) cannot be neglected) will
be considered in a separate paper.

As a consequence, $\Psi_k^{\rm (rad)}(\tau)$ is proportional to
${\cal R}_k^{\rm (rad)}(\tau_i)$, and one has, instead of Eq. (\ref{Pdeltak}), the
simple connection between density perturbation at any time and curvature perturbation at
initial moment of time
\begin{equation} \label{Pdeltak2}
{\cal P}_\delta(k, t) = \left[ \frac{2}{3} (k \tau)^2
\frac{\Psi_k(\tau)}{ {\cal R}_k(\tau_i)} \right]^2 {\cal P}_{\cal R}(k, t_i),
\end{equation}
where the expression for $\Psi_k(\tau)$ is given by \cite{Lyth:2005ze}
\begin{eqnarray}
\label{PsiSol} \Psi_k(\tau) = \frac{2 {\cal R}_k^{\rm (inf)}(\tau_i)}{x^3} [
(x-x_i) \cos(x-x_i) - 
(1+xx_i)\sin(x-x_i) ] .
\end{eqnarray}
\end{widetext}
Correspondingly, in this approximation only one input value, ${\cal R}_k^{\rm (inf)}(\tau_i)$,
is needed for a calculation of the Bardeen potential.

\begin{figure} [!b]
\includegraphics[trim = 0 -10 0 0,  width=0.4 \textwidth]{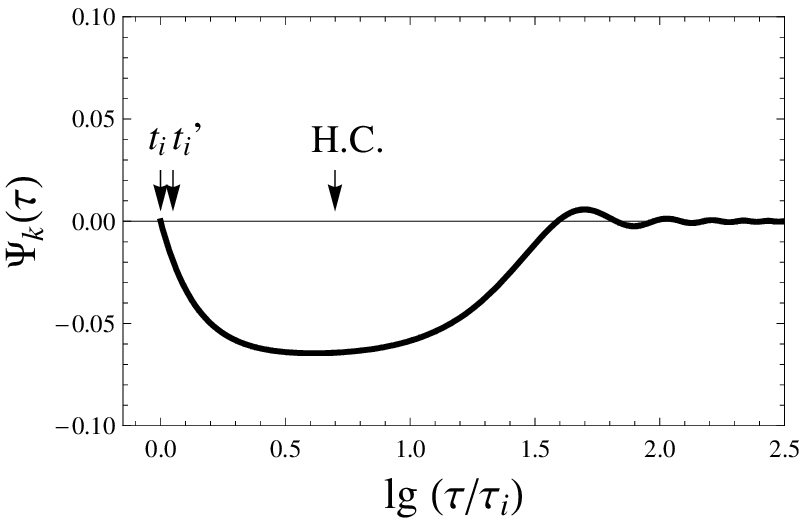} %
\\
\includegraphics[trim = 0 -20 0 0,  width=0.4 \textwidth]{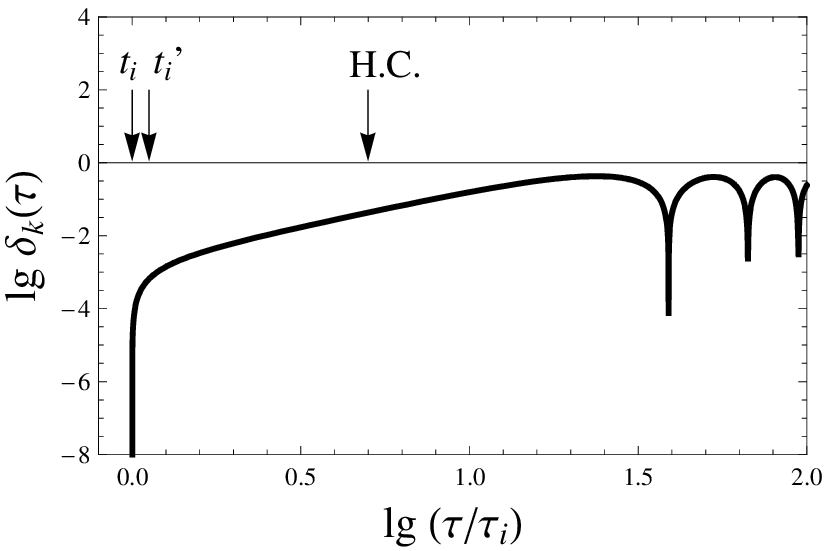} %
\caption{The dependencies of the gravitational potential $\Psi_k$
(upper panel) and the density contrast $\delta_k$ (lower panel) on
time. H.C. denotes the moment of horizon entry for the mode. The wave
number taken is $k=0.2 \tau_i^{-1}$, and normalization is ${\cal R}_k^{\rm (inf)}(\tau_i)
= 0.1$.} \label{psi-fig}
\end{figure}

In Fig. \ref{psi-fig} typical results of the calculation using Eqs. (\ref{PsiSol}) and (\ref{dkt}) are shown.
It is clearly seen from the figure that in our approximate approach
(instantaneous transition from inflation to the radiation era) the evolution of the Fourier transform of the
density contrast starts from zero value. In practical calculations of PBH production we displace the starting
moment from $t_i$ to $t_i'$ ignoring this production inside the short time interval $\Delta t=t_i'-t_i$.
Specifically, the shift $\Delta t$ was determined by the condition $\lg (t_i'/t_i)=0.1$, throughout
all numerical calculations.

\subsection{Power spectrum with maximum}

One of scenarios in which significant PBH production is possible is the one with the
power spectrum ${\cal P}_{\cal R}(k)$ having a strong bump near some value $k=k_0$ in the region of small scales.
Such models have been considered in many papers, including
\cite{Ivanov:1994pa, GarciaBellido:1996qt, Randall:1995dj,  Yamaguchi, Kawasaki, Saito:2008em, Bugaev:2008bi}.

Having the exact form of the power spectrum ${\cal P}_{\cal R}(k)$, it is possible to calculate the PBH mass distribution
using the formalism introduced in the previous sections. However, to be more general, it is convenient to
introduce some parametrization of ${\cal P}_{\cal R}(k)$, and use it to calculate PBH mass spectra.
For example, we find that power spectra with peaks considered in \cite{Bugaev:2008bi} can be rather well parameterized in the region of
the maximum with a distribution of the form
\begin{equation}
\label{PRparam} %
\lg {\cal P}_{\cal R} (k) = B + (\lg {\cal P}_{\cal R}^0 - B)
\exp \Big[-\frac{(\lg k/k_0)^2}{2 \Sigma^2} \Big]
\end{equation}
(it is similar to the primordial spectrum parametrization used in \cite{Green:1999xm}
and in our previous work \cite{Bugaev:2006fe}).
Here, the value of $B$ is known from observations at large scales
($B \approx - 8.6$), and its exact value does not affect the PBH
production rate. ${\cal P}_{\cal R}^0$ is the maximum value
approached by the power spectrum, $k_0$ and $\Sigma$ are
parameters determining the position and width of the peak.

\begin{figure} [!b]
\includegraphics[trim = 0 -30 0 0,  width=0.35 \textwidth]{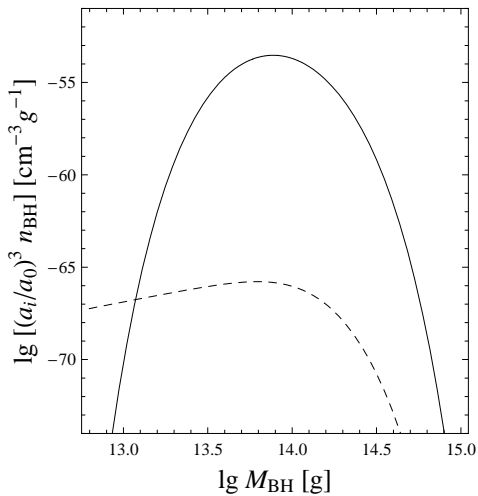} %
\caption{PBH mass spectra calculated for the models of the Carr-Hawking collapse
(solid line) and the critical collapse (dashed line). The following
set of the parameters was used: $\Sigma=3, {\cal P}_{\cal R}^0=0.02, M_h^0=10^{14} {\rm g}.$}
\label{nBHfig}
\end{figure}

It is clear that any power spectrum with maximum can be more or less well approximated in the most
important region (near the maximum) just by three parameters,
determining the position of the peak, its height and width. Thus, the results from taking the spectrum in such a
form will be quite general, because PBH production is mostly determined, of course,
by the region of the spectrum near the maximum.

The connection between the wave number $k_{R}$ and
horizon mass $M_h$ at the moment when the mode enters the horizon
is determined from (\ref{kflaiHi}) and (\ref{MhMiM}) and is
\begin{equation}
\label{MhkflMiaiHi} M_h(k_{R}) = \frac{M_i (a_i H_i)^2}{k_{R}^2}
,
\end{equation}
and further we will use a notation $M_h^0 = M_h(k_0)$, which is the horizon
mass corresponding to a value of $k_0$.

The result of PBH mass spectrum calculations with the particular set
of parameters is shown in Fig. \ref{nBHfig}. It is seen that the mass spectrum strongly depends on the model of
the gravitational collapse (at the same values of the fluctuation
spectrum parameters). One should note that on this and some following
figures we show on the vertical axis the quantity $n_{BH}\times (a_i/a_0)^3$,
which is the comoving number density of PBHs, and is independent
on the reheating temperature $T_{RH}$ in the limit $k_{i} \equiv a_i H_i \gg
k_0$. The case when values of $k \sim k_i$ are important for the PBH
production will be discussed in Sec. \ref{RM-sec}.

\begin{figure} 
\includegraphics[trim = 0 -5 0 0, width=0.34 \textwidth]{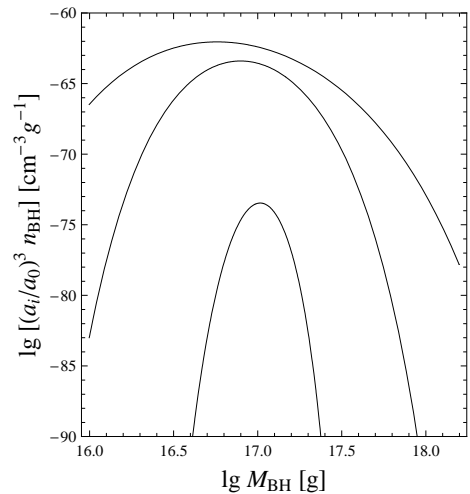} %
\caption{PBH mass spectra for the standard collapse case with ${\cal P}_{\cal R}^0=0.0172, M_h^0=10^{17} {\rm g}$ (for all curves); from bottom to top,
$\Sigma = 1, 3, 5$.} \label{nBH-sig-fig}
\end{figure}

Figure \ref{nBH-sig-fig} shows a dependence of the PBH mass spectrum
on the parameter $\Sigma$ in Eq. (\ref{PRparam}). We can see that this dependence
is rather strong, that comes from the fact that $n_{BH}$ has an
exponential sensitivity to the mass variance $\sigma_R(M)$ (which, in turn, is
determined by the shape of the spectrum, including its width).
We also see that the mass spectrum is always rather strongly peaked near the PBH mass $\sim M_h^0$.

The fraction of an energy density of the Universe contained in PBHs today, $\Omega_{\rm PBH}$,
assuming that a mass of the produced black hole does not change in time, is
\begin{equation}
\label{OmegaPBH}
\Omega_{\rm PBH} = \frac{1}{\rho_c} \left(
\frac{a_i}{a_0}\right)^3 \int M_{BH} n_{BH}(M_{BH}) d M_{BH}
\end{equation}
($\rho_c$ is the critical density). This formula is rather accurate for black holes with
initial mass $M_{BH} \gg M_*$, where $M_* \approx (3 t_0 \alpha_0)^{1/3} \approx 5\times 10^{14}$ g
is the initial mass of PBH which reaches its final state of evaporation
today \cite{Page:1976df}, $\alpha_0=8.42\times 10^{25}\; {\rm g}^3 {\rm s}^{-1}$,
and $t_0$ is the age of the Universe.

One should note that for the concrete PBH mass spectra shown in Fig. \ref{nBH-sig-fig}, we have, for the case of $\Sigma=5$, the inequality $\Omega_{PBH} > 1$,
which is an inconsistent result, so the production of such a big amount of PBHs is forbidden.
We will discuss the possibility of a constraining the model parameters in the distribution
of Eq. (\ref{PRparam}) in detail further, in Sec. \ref{sec-constr-peak}.
The value ${\cal P}_{\cal R}^0=0.0172$ chosen for the calculation of the spectra shown in Fig. \ref{nBH-sig-fig}
corresponds, in the case of $\Sigma=3$, to the value  $\Omega_{PBH}$ equal to $\approx 0.23$.
This is in the range needed to explain the observed amount of non-baryonic dark matter \cite{Amsler:2008zz},
$\Omega_{\rm nbm} h^2 = 0.106 \pm 0.008$ (with $1\sigma$ uncertainty).
For the case of $\Sigma=1$, it appears that $\Omega_{PBH} \sim 10^{-10}$, probably
too small of a fraction to produce observable cosmological consequences.

\section{Neutrino and photon spectra from PBHs evaporations}
\label{sec-nu-gamma}

PBH evaporation, predicted by Hawking \cite{Hawking:1974sw}, causes evaporating black holes to produce an isotropic
extragalactic photon and neutrino backgrounds, which can be, at least in
principle, measured experimentally.
It can be calculated having the PBH mass spectrum, and the comparison with observations can be made.

Evolution of a PBH mass spectrum due to the evaporation leads to
the approximate expression for this spectrum at any moment of
time:
\begin{equation}
\label{nBHmt} n_{BH}(m,t)=\frac{m^2 }{(3\alpha t + m^3)^{2/3}}
n_{BH}\left((3\alpha t + m^3)^{1/3}\right),
\end{equation}
where $\alpha$ accounts for the degrees of freedom of evaporated
particles and, strictly speaking, is a function of a running value
of the PBH mass $m$. In our numerical calculations we use the
approximation
\begin{equation}
\label{alphaconst} \alpha={\rm const}=\alpha (M_{BH}^{max}),
\end{equation}
where $M_{BH}^{max}$ is the value of $M_{BH}$ in the initial mass
spectrum corresponding to a maximum of this spectrum. Special
study shows that errors connected with such an approximation are
rather small. In this work we use parametrization of the function $\alpha(M_{BH})$
presented in \cite{BugaevD65}.

The expression for the extragalactic differential energy spectrum
of neutrinos or photons (the total contribution of all black
holes) integrated over time is \cite{BugaevD65}
\begin{widetext}
\begin{eqnarray}
\label{SE} S(E)= \frac{c}{4\pi} \int dt
\frac{a_0}{a}\left(\frac{a_i}{a_0}\right)^{3} \int n_{BH}
\left((3\alpha t + m^3)^{1/3}\right)
\cdot \varphi(E (1+z),m) e^{-\tau(E,z)} \frac{m^2 dm}{(3\alpha t + m^3)^{2/3}}  \equiv \nonumber\\
\\
\equiv \int F(E,z)d \lg (z+1).  \nonumber
\end{eqnarray}
\end{widetext}

In this formula, $a_i$, $a$, and $a_0$ are cosmic scale factors at
$t_i$, $t$ and at present time, respectively, and $\varphi(E,m)$
is a total instantaneous spectrum of the radiation (neutrinos or
photons) from an evaporation of the individual black hole. The
exponential factor in Eq. (\ref{SE}) takes into account an
absorption of the radiation during its propagation in space. The
processes of neutrino absorption are considered, in a given
context, in \cite{BugaevD65}.
In the last line of (\ref{SE}) we changed the variable $t$ on
red shift $z$ using the flat cosmological model with $\Omega_{\Lambda}\ne 0$
for which
\begin{eqnarray}
\label{dtdz} \left|\frac{dt}{dz}\right|=\frac{1}{H_0} \frac{ \left(\Omega_M
(1+z)^{3}+\Omega_R (1+z)^{4} + \Omega_\Lambda\right)
^{-1/2} } {(1+z)} 
\end{eqnarray}
($H_0=100 h \; {\rm km/(s\cdot Mpc)}$, $\Omega_R = \Omega_\gamma + \Omega_\nu$),
with the set of the basic cosmological parameters taken from \cite{Amsler:2008zz}:
$h=0.73$, $\Omega_M h^2= \Omega_{nbm} h^2 + \Omega_B h^2 =0.128$,
$\Omega_\gamma h^2 = 2.47\times 10^{-5}$, $\Omega_{\Lambda}=1-\Omega_M-\Omega_R$.
The neutrino fraction in the total energy density, $\Omega_\nu$, is calculated assuming that the neutrino masses
are negligible, in this case $\Omega_\nu=\frac{7}{8} (\frac{4}{11})^{4/3} \times 3
\Omega_\gamma \approx 0.68 \Omega_\gamma$  (the term with $\Omega_R$ is
essential in Eq. (\ref{dtdz}) only for $z \gtrsim \Omega_M/ \Omega_R$).


\begin{figure} [!b]
\includegraphics[trim = 0 0 0 -0, width=0.4 \textwidth]{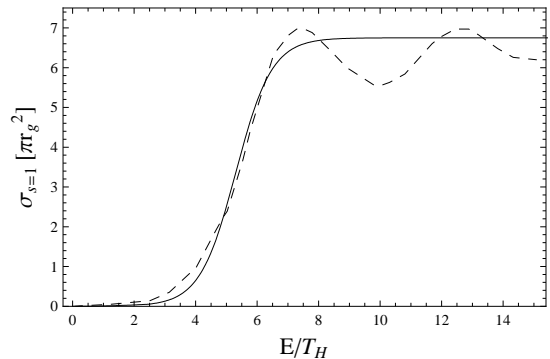} %
\caption{Absorption cross section $\sigma_s$ for photons, in units of $\pi r_g^2$.
Dashed line is the result of the numerical calculation \cite{MacGibbon:1990zk},
solid line is the approximation (\ref{gaga}). }
\label{sig-ga-fig}
\end{figure}

The direct Hawking flux of evaporated particles (per degree of freedom) is
\begin{equation}
\varphi(E,M_{BH}) = \frac{dN}{dE dt} = \frac{\Gamma_s}{2 \pi} \frac{1}{\exp(E/T_H) - (-1)^{2s}},
\end{equation}
where $\Gamma_s$ is a grey-body factor ($\Gamma_s=\sigma_s E^2 / \pi$ for massless particles, where
$\sigma_s$ is the absorption cross section), and $T_H$ is the Hawking temperature
\begin{equation}
T_H = \frac{m_{Pl}^2}{8 \pi M_{BH} } = 1.06 \left( \frac{10^{13} {\rm g} }{M_{BH} } \right) {\rm GeV}.
\end{equation}
In the high energy limit ($E \gg T_H$), the cross section $\sigma_s$ approaches a constant value independent of $s$,
and $\Gamma_s$ in this case is
\begin{equation}
\Gamma_s = \frac{27}{64 \pi^2} \frac{E^2}{T_H^2}.
\end{equation}
For lower energies, however, this factor needs to be calculated numerically.

For photons, we use the following approximate parametrization of this function \cite{KlimaiThesis}
\begin{eqnarray}
\Gamma_\gamma \approx %
\left\{  \begin{array} {l}
  \frac{27}{64 \pi^2} \frac{E_\gamma^2}{T_H^2} \left( \exp[9.08-1.71(E_\gamma/T_H)] +1 \right)^{-1},
  \\ \;\;\;\;\;\;\;\; \;\;\;\;\;\;\;\; \;\;\;\;\;\;\;\; \;\;\;\;\;\;\; \;\;\;\;\;\;\;\;\;\;  E_\gamma \ge 2.5 T_H ,\\
  \frac{4}{3} \left( \frac{E_\gamma}{4 \pi T_H} \right)^4 , \;\; E_\gamma < 2.5 T_H \; .
\end{array} \right.
\label{gaga}
\end{eqnarray}
Here, the low-energy part comes from analytically obtained relation
$\sigma_{s=1}(E\to 0) = \frac{4}{3}\pi r_g^2 (E r_g)^2$
\cite{Page:1976df} ($r_g$
is the Schwarzschild radius). The higher energy part is just an approximation smoothing the oscillations
in $\sigma_s$, see Fig. \ref{sig-ga-fig}. The form of Eq. (\ref{gaga}) at high energies is
analogous to one used in \cite{Daghigh:2002fn} for neutrinos,
\begin{equation}
\Gamma_\nu = \frac{27}{64 \pi^2} \frac{E^2}{T_H^2} \left(0.075 + \frac{0.925}{\exp[5-1.607(E/T_H)]+1}  \right).
\end{equation}

The instantaneous spectra of neutrinos and photons from PBH
evaporations were calculated using the model of Ref. \cite{MacGibbon:1990zk}
and parametrization of Ref. \cite{Halzen:1991uw}.
In these papers the particles directly emitted from the black hole (through the Hawking mechanism) as well
as those produced in processes of quark fragmentations and subsequent decays of mesons had been
taken into account.

\begin{figure} 
\includegraphics[trim = 0 -12 0 0, width=0.41 \textwidth]{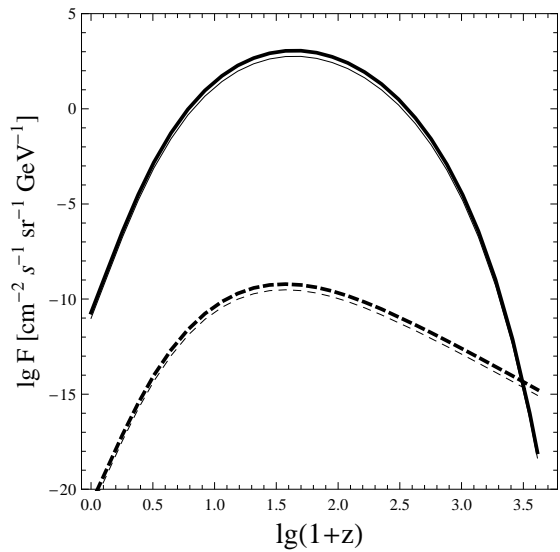} %
\caption{Red shift distribution of the integrand $F(E,z)$.
Thick lines are for neutrinos, thin lines are for photons (absorption of
$\gamma$-rays at $z\gtrsim 700$ is not shown in this figure).
Dashed curves represent the case of the critical collapse, solid ones correspond to the standard collapse.
PBH mass spectra shown in Fig. \ref{nBHfig} were used in the calculation, $E=1$GeV for all cases. }
\label{Ffig}
\end{figure}

In Fig. \ref{Ffig} the red shift distributions of the differential
energy spectra $S(E)$ are shown for the neutrino or photon with energy
1 GeV. One can see that the distributions are more wide in the
case of the critical collapse (because the PBH mass spectrum in
this case has a long "tail" of small masses). The absorption
factor $e^{-\tau}$ is efficient: in case of photons, the maximum value of $z$ for
which the absorption can be neglected is around $700$
\cite{Zdz}; for neutrinos the absorption is important at $z \gtrsim 10^6
\div 10^7$ \cite{BugaevD65}.

\section{Constraints on the power spectrum with maximum}
\label{sec-constr-peak}

Assuming that ${\cal P}_{\cal R}(k)$ has the form of Eq. (\ref{PRparam}), we can calculate the produced PBH
mass spectra, compare the possible consequences of the existence of such amount of PBHs with
observations, and put limits on the parameters of this distribution.
The calculation of the actual constraints in this paper (as well as in our
previous work \cite{Bugaev:2006fe}) uses the following basic observational
facts.

1. The differential energy spectrum of the extragalactic photon
background in the wide region of energies is known
\cite{Strong:2004ry}. For our purpose, the most interesting range
is about $E_{\gamma} \sim 1$ MeV $\div 1$ GeV, and the order of
magnitude of the flux is roughly given by $E^2 S_\gamma(E) \sim 10^{-6} $
GeV cm$^{-2} $ s$^{-1}$ sr$^{-1}$ in this region. We obtain our
constraints just from the condition that the diffuse gamma ray
flux produced by PBHs does not exceed the observed extragalactic
one.

2. According to the data of Super-Kamiokande experiment
\cite{Malek:2002ns}, the electron antineutrino background flux in extragalactic
space is constrained by the inequality
\begin{equation}
\label{nuconstr} \Phi(E_{\tilde \nu_e}>19.3 \; {\rm MeV}) < 1.2 \;
{\rm cm}^{-2} {\rm s}^{-1}.
\end{equation}
We integrate the calculated flux and put constraints from the condition that it does not
exceed this limit. We note, however, that the limit (\ref{nuconstr}) is not a model-independent one.
It was obtained assuming exponentially falling neutrino spectrum, which is an expectation for
the extragalactic diffuse flux from supernovae. The neutrino
flux from PBHs in the mass region we consider is falling as
$E_\nu^{-3}$, which, in fact, will change the constraint (\ref{nuconstr}). Estimations show, however, that this change
is about a factor of 2, and it is not very significant for putting constraints on such parameters
as ${\cal P}_{\cal R}^0$. Similar uncertainty also comes from the inclusion of neutrino oscillations
in the analysis, which we do not take into account.

3. The fraction of the energy density of the Universe contained in PBHs, $\Omega_{\rm PBH}$,
which can be calculated using formula (\ref{OmegaPBH}),
cannot exceed the one for non-baryonic dark matter, $\Omega_{\rm nbm}$ \cite{Amsler:2008zz}.
This constraint is important for black holes with
initial mass $M_{BH} > M_*$, i.e., ones that did not evaporate up to the present time.

The actual condition we use is $\Omega_{PBH} < 0.3$ (having all the uncertainties considered,
$\Omega_{\rm nbm}=0.3$ is still allowed by the observations within the $3\sigma$ confidence interval,
however, the uncertainty in this number does not seriously affect the power spectrum constraint).

\begin{figure}
\includegraphics[trim = 0 -2 0 0, width=0.44 \textwidth]{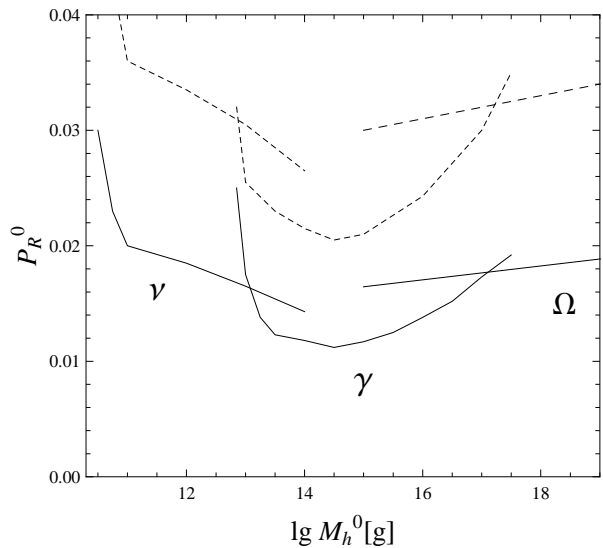} %
\caption{Constraints on the parameter ${\cal P}_{\cal R}^0$
(giving the height of the peak in Eq. (\ref{PRparam})), that
were obtained in this work (for this figure, we used $\Sigma=3$).
The forbidden values of the parameter ${\cal P}_{\cal R}^0$
lie above the curves shown in the figure.
Constraints following
from the neutrino and the gamma ray experiments are shown together with
the energy density constraint. Dashed lines correspond to the model of
the critical collapse, solid lines represent the results obtained
using the standard collapse picture.
} \label{const}
\end{figure}

The resulting constraints obtained from all the conditions discussed
are shown in Fig. \ref{const}.  The value of the
parameter $\Sigma$ characterizing the width of the gaussian
distribution in Eq. (\ref{PRparam}) was fixed throughout all
calculations ($\Sigma = 3$). The constraints are given as a
function of the horizon mass $M_h^0$. The connection of $M_h^0$ with the corresponding wave number
$k_0$ is very simple and is given by the formula following from Eq. (\ref{MhkflMiaiHi}):

\begin{equation}
k_0 \approx \frac{3 \times 10^{23}} {\sqrt{M_h^0 / 1 {\rm g} } } {\rm Mpc^{-1} } .
\end{equation}

It follows from these results that the constraints are stronger under assumption of the
standard Carr-Hawking collapse, which is expected because in this case the
threshold density contrast leading to PBH formation is smaller.
The constraints are independent on the reheating temperature if it
is high enough to provide the condition $M_i \ll M_h^0$ (this is true for all
PBH masses considered if $T_{RH} \gtrsim 10^{11}$ GeV).

The second important result is that the constraints based on the
neutrino emission of PBHs are comparable with those following from the
photon emission and from the gravitational constraint. At the region of small horizon masses (large
$k_0$), where large red shifts are important, the constraints
from the neutrino emission are stronger.

Fig. \ref{const} is analogous to one presented in our previous work \cite{Bugaev:2006fe}.
In the present paper we add the gravitational constraint on $\Omega_{PBH}$ and more carefully investigate
the mass region $M_h^0\sim 10^{16} \div 10^{17}$ g. In this region, constraints from
energy density and from gamma ray background coexist. We do not extend in the present paper the neutrino constraints
to masses $M_h^0 \gtrsim M_*$ because the neutrino background in this case strongly deviates from the simple
behavior $\sim E^{-3}_\nu$ and additional analysis is needed to estimate how the limit of Eq. (\ref{nuconstr})
will change in such a situation.

\section{Constraints on the running mass model from PBHs}
\label{RM-sec}

The running mass inflation model was proposed in \cite{Stewart:1996ey, Stewart:1997wg} and further
studied in many papers including \cite{Covi:1998jp, Covi:1998mb, Covi:1998yr, German:1999gi,
Covi:2004tp}. The model predicts a rather strong scale dependence of the spectral index, possibly
allowing large values of ${\cal P}_{\cal R}(k)$ at small scales, which can lead to PBH production
[\cite{Leach:2000ea}, \cite{BugaevD66}, \cite{Bugaev:2008bi}]. The first constraints on the running mass model parameters from PBHs  were obtained in works \cite{Leach:2000ea, BugaevD66}. In previous works the PBH mass spectrum in the running mass model have not been
calculated in detail. We do it in the present paper to demonstrate how the general formalism developed in
Sec. \ref{sec-form} works.

We will also show how it is possible to constrain such
observable parameters as the spectral index measured at
cosmological scales, $n_0$, and its running $n'_0$, assuming the
running mass model is correct.

\begin{figure} [!b]
\includegraphics[trim = 0 -8 0 0, width=0.48 \textwidth]{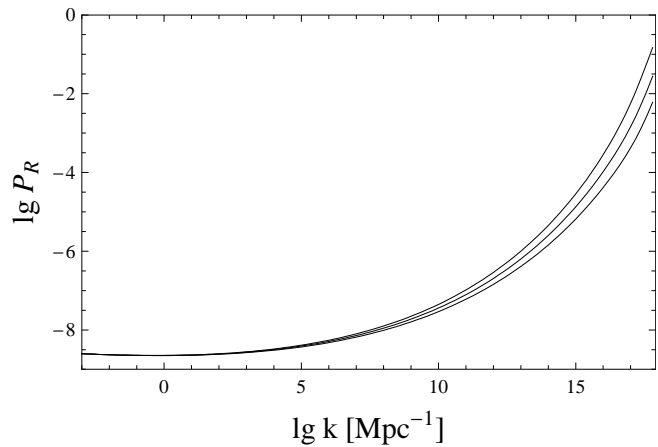} %
\caption{Power spectrum ${\cal P}_{\cal R}(k)$, calculated for the
running mass model, with $T_{RH}=10^{10}$ GeV, $n_0=0.97$, from
top to bottom, $10^3 n_0'=4.3, 4.5, 4.7$ .} \label{PR-rm-fig}
\end{figure}

The potential of the running mass model takes into account quantum corrections in the context of the softly
broken global supersymmetry and is given by the formula
\begin{equation}
V = V_0 + \frac{1}{2} m^2(\ln \phi) \phi^2 \;.
\end{equation}
The dependence of the inflaton mass $m(\ln \phi)$ on renormalization scale $\phi$ is
determined by the renormalization group equation (RGE).

\begin{figure} 
\includegraphics[trim = 0 0 0 0, width=0.34 \textwidth]{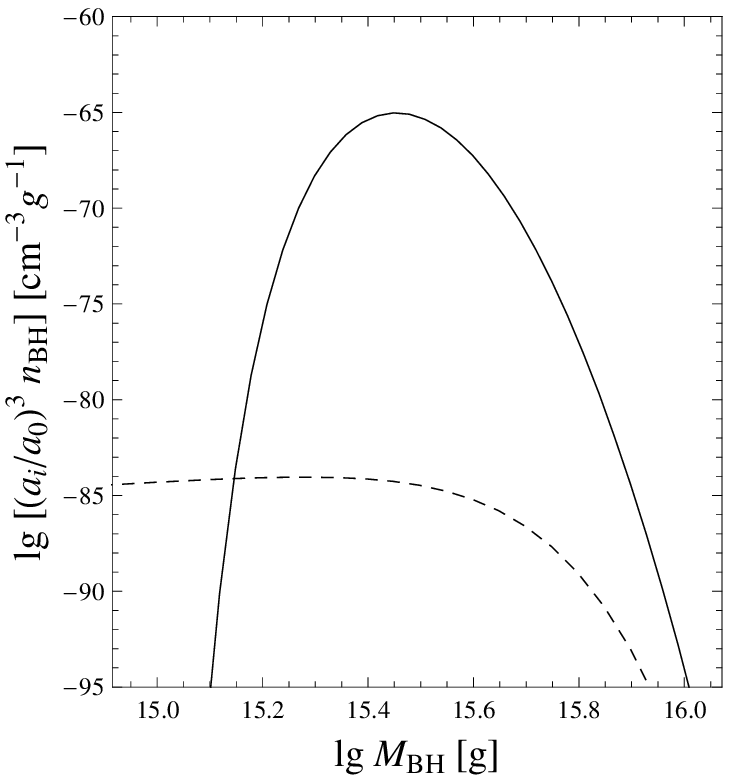} %
\\
\includegraphics[trim = 0 0 0 0, width=0.34 \textwidth]{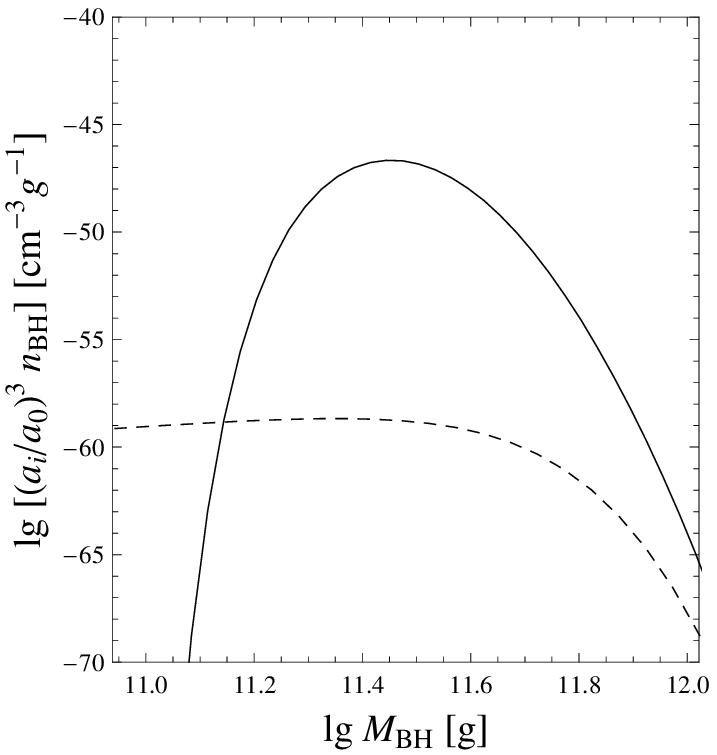} %
\caption{PBH mass spectra in the running mass model, calculated for the following set of parameters:
$T_{RH}=10^8$ GeV, $n_0=0.96$, $n_0'=5.33 \times 10^{-3}$ (upper panel);
$T_{RH}=10^{10}$ GeV, $n_0=0.96$, $n_0'=4.4 \times 10^{-3}$ (lower panel).
Solid line represents the result for
the standard collapse case, dashed line corresponds to the critical collapse.
} \label{nBH-rm-fig}
\end{figure}

Apart from $V_0$, the shape of the running mass
potential is determined by parameters $c$ and $s$, which are connected to the
other quantities by
\begin{eqnarray}
c \frac{V_0}{M_P^2} = - \left. \frac{d m^2}{d \ln \phi} \right|_{\phi=\phi_0} \; , \\
s = \frac{H_I}{2 \pi \phi_0 {\cal P}_{\cal R}^{1/2}(k_0) } \;\;\;\;\;\;\; \;\;
\end{eqnarray}
($H_I^2 = V_0/3 M_P^2$, and $\phi_0$ is the value of the inflaton field corresponding to the
cosmological scale $k_0$). They are also linked to observables $n_0$ and $n_0'$:
\begin{equation}
n_0-1 \approx 2 (s-c) \;\; , \;\; n_0' \approx 2 s c \;,
\end{equation}
and we see that measuring $n_0$ and $n_0'$ allows us to reconstruct the potential shape in this model,
and the only free parameter left is $V_0$ (or $H_I$). From the theoretical point of view \cite{Covi:2004tp},
$H_I$ can lie in the wide range of values from, say, $H_I\sim 10^4$ GeV for anomaly-mediation case to $H_I \sim 10^{-3}$ GeV
for gauge-mediation. Assuming instant reheating at some value of the inflaton
field, from the energy conservation condition we have:
\begin{equation}
\frac{\pi^2}{30} g_* T_{RH}^4 = V_0
\end{equation}
($g_* \sim 100$), and the reheat temperature turns out to be in the region $10^7 \div 10^{11}$ GeV
for the given values of $H_I$. For further calculation we will use two values of $T_{RH}$, namely,
$10^8$ and $10^{10}$ GeV.

It was noted in \cite{Bugaev:2008bi}, that for the running mass
model, the numerical calculation of the power spectrum is necessary in
the region of large $k$ values. In this work we also use the numerical
calculation of power spectra, not relying on the analytical
approximations. Several results for ${\cal P}_{\cal R}(k)$
are given in Fig. \ref{PR-rm-fig}, which shows how
sensitive is the value of ${\cal P}_{\cal R}$ near the end of
inflation to the change in $n_0'$.

Some examples of the PBH mass spectrum calculation are given in Fig. \ref{nBH-rm-fig} for
two models of the collapse considered above. The order of magnitude of the PBH mass produced
is equal to the initial horizon mass $M_i$, which is given by
\begin{equation}
M_i = \frac{4}{3}\pi t_i^3 \rho_i \approx 0.038 \frac {m_{Pl}^3} {g_*^{1/2} T_{RH}^2}.
\end{equation}
The real PBH mass spectrum, and, particularly, its maximum, is, however, determined by the dynamics of their formation, including
the dependence of the gravitational potential $\Psi_k$ on
time. This dependence is very important in the region of smallest possible PBH
masses produced, i.e., in the case $k_{R} \sim k_i$.
For example, in the model of the standard collapse with a monotonous
power spectrum we would expect most of PBHs to be produced near
the smallest possible mass, i.e., near $M_{BH} \approx \sqrt{1/3}\;
M_i$ (e.g., $M_{BH} \approx 7\times 10^{14}$ g for $T_{RH}=10^8$ GeV).
From Fig. \ref{nBH-rm-fig} we see, however, that the actual
maximum is near $M_{BH} \approx 3\times 10^{15}$ g in this case, which is due
to suppression of PBH production at smaller masses.

\begin{figure}
\includegraphics[trim = 0 0 0 0, width=0.43 \textwidth]{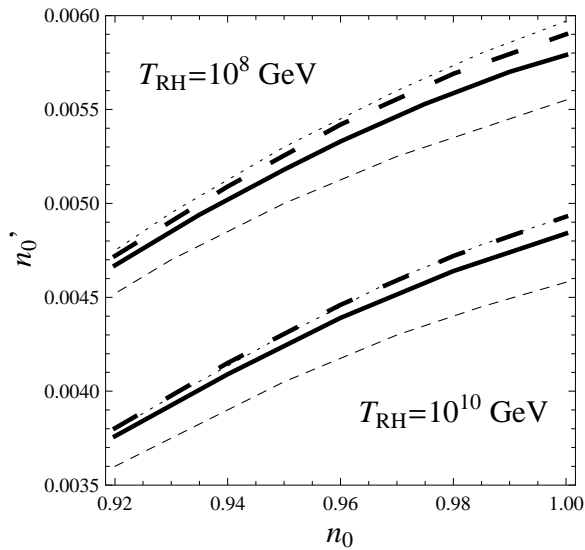} %
\caption{Constraints on the running mass model coming from PBHs, for two values of parameter $T_{RH}$.
The forbidden regions of parameters lie above the corresponding curves.
Solid thick lines are constraints for the standard collapse model,
dashed thick lines are those for the critical collapse.
Thin lines show, for a comparison, the values of $n_0$ and $n_0'$, for which ${\cal P}_{\cal R}=10^{-1}$ (dotted line)
and ${\cal P}_{\cal R}=10^{-2}$ (dashed line) at maximum value of $k$, $k=k_i$.
} \label{constr-rm-fig}
\end{figure}

Fig. \ref{constr-rm-fig} shows the constraints on $(n_0, n_0')$ obtained assuming
that the running mass model is correct. Reconstructing the potential from values of $n_0$, $n_0'$
and $T_{RH}$, we calculate the power spectrum ${\cal P}_{\cal R}(k)$ in the whole region of
wave numbers up to $k_i$, and the corresponding PBH mass spectrum.
The method of obtaining the constraints depends on the characteristic PBH mass produced:
for $T_{RH}=10^{10}$ GeV, the PBH mass in maximum of the distribution is about
$3\times 10^{11}$ g (see Fig. \ref{nBH-rm-fig}), and neutrino background data was used to
constrain such models, because, as we have seen in Sec. \ref{sec-constr-peak}, neutrino
experiments are adequate to constrain PBHs of such initial masses.
For $T_{RH}=10^{8}$ GeV, the PBH mass in maximum of the spectrum is about $3\times
10^{15}$ g, and constraints are
obtained using the extragalactic gamma ray background.

We see from Fig. \ref{constr-rm-fig} that, assuming running mass model is really responsible for
inflation and production of perturbation spectrum, running of the spectral index at cosmological scales
is strongly constrained by PBH overproduction and cannot be larger than $(3 \div 6) \times 10^{-3}$
(the exact value of the constraint depends on other parameters).

\section{Dark matter and PBHs}
\label{sec-DM}

PBHs produced in the early Universe, if they are heavy enough not to evaporate till the present day,
can contribute to dark matter, and potentially it is possible to explain all the
non-baryonic dark matter with them. For example, the mass spectrum shown in Fig. \ref{nBH-sig-fig} (for the case
of $\Sigma=3$) gives $\Omega_{PBH} \approx 0.23$.

In works \cite{Frampton:2005fk, Bambi:2008kx}, PBHs clustered in the Galactic bulge are proposed to explain the
observed $511$ keV photon line from the
Galactic center \cite{Knodlseder:2005yq, Jean:2005af}. Positrons evaporated from PBHs annihilate with electrons
from the interstellar medium thus producing the observed photons. Assuming that Solar System lies at a distance of
$r_0 = 8.5$ kpc from the Galactic center, the total positron production rate needs to be about
$3\times 10^{43}$ s$^{-1}$, and the corresponding characteristic value of the PBH mass should be in the range
$\sim 10^{16} \div 10^{17}$ g (this was shown in ref. \cite{Bambi:2008kx} from calculations of the
photon background in extragalactic space and the photon flux from the Galactic center from PBH evaporations).
We will show in this paper that our distribution presented in Fig. \ref{nBH-sig-fig} (for $\Sigma=3$)
is also capable to explain the needed positron production rate.

\begin{figure} [!b]
\includegraphics[trim = 0 0 0 -0, width=0.48 \textwidth]{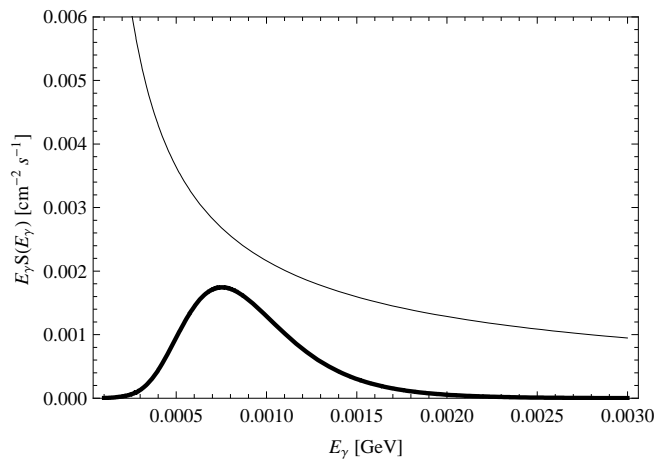} %
\caption{Gamma ray flux from PBHs clustered in the Galactic center, assuming
their total mass is $10^9 M_\odot$
and mass distribution is the same as in Fig \ref{nBH-sig-fig} ($\Sigma=3$) (thick line).
The thin line shows the measured background flux.}
\label{fromGC-fig}
\end{figure}

Assuming that the PBH mass distribution in Galaxy does not change compared to the initial one,
for the PBH mass distribution in the Galactic center $N_{BH}$ we can write a simple proportionality relation
\begin{equation}
N_{BH}(M_{BH}) = C \; n_{BH}(M_{BH}),
\end{equation}
where $C$ will be determined by the total mass of PBHs:
\begin{equation}
\int N_{BH}(M_{BH}) M_{BH} d M_{BH} = M_{tot}.
\end{equation}
The gamma ray flux on the Earth surface produced by such PBHs is
\begin{eqnarray}
\label{flux-earth}
\frac{dN_\gamma}{dE_\gamma dS dt} =
\;\;\;\;\;\;\;\;\;\;\;\;\;\;\;\;\;\;\;\;\;\;\;\;\;\;\;\;\;
\;\;\;\;\;\;\;\;\;\;\;\;\;\;\;\;\;\;\;\;\;\;\;\;\;\;\;\;\;\;\;\;\;
\\ = \nonumber \frac{1}{4\pi r_0^2}
\int N_{BH}(M_{BH}) \frac{dN_\gamma}{dE_\gamma dt}(M_{BH}, E_\gamma) dM_{BH},
\end{eqnarray}
and the positron production rate in the Galactic center is
\begin{eqnarray}
\frac{dN_{e^+} }{dt} =
\;\;\;\;\;\;\;\;\;\;\;\;\;\;\;\;\;\;\;\;\;\;\;\;\;\;\;\;\;
\;\;\;\;\;\;\;\;\;\;\;\;\;\;\;\;\;\;\;\;\;\;\;\;\;\;\;\;\;\;\;\;\;
\;\;\;\;\;\;\;\;\;
\\ = \nonumber
\int\int N_{BH}(M_{BH}) \frac{dN_{e^+} }{dE_{e^+} dt}(M_{BH}, E_{e^+}) dM_{BH} dE_{e^+}.
\end{eqnarray}

The total mass needed can be expressed through the observable positron production rate and the initial
mass spectrum by the formula
\begin{equation}
M_{tot} = \frac
{ \frac{dN_{e^+} }{dt} \times \int n_{BH}(M_{BH}) M_{BH} dM_{BH}  }
{ \int n_{BH}(M_{BH}) \frac{dN_{e^+} }{dE_{e^+} dt}(M_{BH}, E_{e^+}) dE_{e^+} dM_{BH} }.
\end{equation}
In our case, we obtain $M_{tot} \approx 10^9 M_\odot$. If these black holes
are uniformly distributed in the spherical region
of radius $r \sim 600$ pc, the clusterization factor needed for that is about
\begin{equation}
\zeta \approx \frac{M_{tot} }{  \frac{4}{3} \pi r^3 \; \Omega_{PBH} \rho_c}  \approx 3\times 10^7 \; .
\end{equation}

The gamma ray flux from such PBHs calculated using Eq. (\ref{flux-earth}) is presented in Fig. \ref{fromGC-fig},
together with the measured continuum, given by the approximate formula \cite{Jean:2005af, Bambi:2008kx}
\begin{equation}
\frac{d\Phi_{\rm cont}}{dE_\gamma} \approx 7 \; {\rm GeV}^{-1} {\rm cm}^{-2}  {\rm s}^{-1}
\left( \frac{E_\gamma}{ 511 \; {\rm keV} } \right)^{-1.75} \; ,
\end{equation}
and we see that the flux from PBHs does not exceed the observational one.

\begin{figure} [!t]
\includegraphics[trim = 0 -12 0 0, width=0.48 \textwidth]{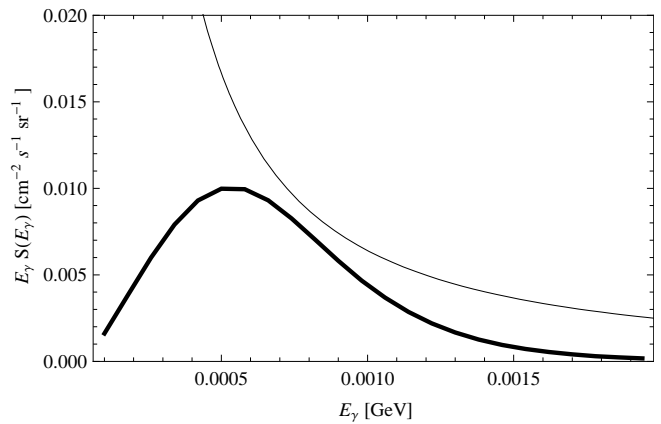} %
\caption{Diffuse gamma ray flux produced by PBHs with the mass spectrum shown in Fig
\ref{nBH-sig-fig} ($\Sigma=3$) (thick line) and the measured extragalactic background
(thin line).}
\label{diffuse-gam-fig}
\end{figure}

For the calculation of diffuse extragalactic photon flux, clusterization of PBHs is not important,
because this flux is sensitive to the average number of PBHs and their mass distribution only.
Thus, formula (\ref{SE}) is suitable for
the calculation. It can, however, be much simplified because such PBHs (with mass sufficiently
larger than $M_*$) do not lose the significant part of their mass in course of the evaporation, and approximation
$M_{BH}(t) \approx {\rm const}$ is adequate, so we can neglect the quantity $3 \alpha t$ in it.
The absorption factor can also be dropped because the effectively working red shifts in this task are small.
The diffuse gamma ray flux from our PBH mass distribution (Fig. \ref{nBH-sig-fig}, $\Sigma=3$) is shown in Fig. \ref{diffuse-gam-fig}.
We see that, again, the experimentally measured flux, given by \cite{Strong:2004ry, Bambi:2008kx}
\begin{equation}
\frac{d\Phi_{\rm extra}}{dE_\gamma} \approx 6.4 \; {\rm GeV}^{-1} {\rm cm}^{-2} {\rm s}^{-1} {\rm sr}^{-1}
\left( \frac{E_\gamma}{ 1 \; {\rm MeV} } \right)^{-2.38} \; ,
\end{equation}
is not exceeded.

\section{Conclusions}
\label{sec-concl}

{\bf 1.} We have developed a method of PBH mass spectrum calculation, which is based on the Press-Schechter
formalism and takes into account the dependence of the gravitational potential on time. We found
that it is more convenient and natural to use in the formula for the mass variance (Eq. (\ref{SigR})) the
gravitational potential $\Psi_k(t)$ calculated as a function of time rather than the
transfer function $T(k,t)$ (the latter is defined, e.g., in \cite{Blais:2002gw}). To derive the PBH mass
spectrum, the power spectrum of curvature perturbations calculated (using inflationary models)
at the end of inflation is needed as an input.

{\bf 2.} PBH mass spectra are calculated in two inflationary models predicting large amplitudes of the curvature
perturbation spectrum at some wave number $k$, namely, in the model with the power spectrum with a peak and
in the running mass model. Examples of such a calculation are given in Figs. \ref{nBHfig}, \ref{nBH-sig-fig} and \ref{nBH-rm-fig}.

{\bf 3.} Constraints on parameters of both these models are obtained using available data on the
extragalactic photon diffuse background and the upper limit on the extragalactic neutrino background,
obtained in the Super-Kamiokande neutrino experiment.

The results are shown in Figs. \ref{const} and
\ref{constr-rm-fig}. It is seen that for the power spectrum with a peak,
the height and the width of the peak can be constrained in rather wide region of wave numbers determining
its position in the spectrum. For the running mass model we obtain the constraint on
the spectral index running at cosmological scales, $n_0' \lesssim (3 \div 6) \times 10^{-3}$
(the exact constraint value depends on $n_0$ and $T_{RH}$).

{\bf 4.} It follows from Fig. \ref{const} that constraints obtained from neutrino
experiments are stronger than those obtained from extragalactic photon background, in the
region of horizon masses from $\sim 10^{11}$ to $\sim 10^{13}$ g. So, constraints from neutrino experiments and from
measurements of diffuse photon background are complementary.

{\bf 5.} PBHs can be the main contributor to the dark matter in the Universe.
It is shown that if, in this case, the PBH mass distribution is peaked near $\sim 8 \times 10^{16}$ g,
as in Fig. \ref{nBH-sig-fig} with $\Sigma=3$,
the experimentally observed galactic $511$ keV photon line can, in fact, be caused by annihilating positrons
produced by evaporating PBHs clustered in the Galactic center. The total PBH mass needed for this to happen
must be about $10^9 M_\odot$, and clusterization factor needed is $\zeta \sim 3\times 10^7$.

\bigskip
\begin{acknowledgments}
The authors are grateful to Professor A.A. Starobinsky for the important remark.
The work was supported by Russian Foundation for Basic Research (Grant No. 06-02-16135).
\end{acknowledgments}

\appendix
\section{Formation of PBHs in a radiation-dominated universe}
\label{app1}

In this Appendix we give basic formulas of the PBH formation process following closely the
work \cite{Bullock:1996at}.

Evolution of a spherically symmetric region with density $\tilde\rho$ greater than
the background density $\rho$ is governed by the positive curvature Friedmann equation,
\begin{equation}
\tilde H^2 (\tilde t) = \frac{8 \pi}{3 m_{Pl}^2} \tilde\rho(\tilde t) - \frac{1}{\tilde {\bf R}^2(\tilde t) }.
\end{equation}
Here, $\tilde H$, $\tilde \rho$ and $\tilde {\bf R}$ are Hubble parameter, density and a physical size of
the region, respectively. In a background space the evolution equation is, correspondingly,
\begin{equation}
H^2(t) = \frac{8 \pi}{3 m_{Pl}^2} \rho(t).
\end{equation}

An expansion of the overdensed region goes on according to the law \cite{Harrison:1969fb}
\begin{equation}
\tilde {\bf R} \sim \tilde t^{1/2},
\end{equation}
and stops at time $t_c$. This time is determined from the condition $\tilde H(t_c) = 0$ and
is given by the expressions
\begin{equation}
t_c = \delta_i^{-1} t_i \;\;, \;\; \delta_i = \frac{\tilde \rho_i-\rho_i}{\rho_i},
\label{tcrho}
\end{equation}
where $t_i$ is the initial moment of time (which is chosen in such a
way that $\tilde H(t_i) = H(t_i) = H_i$ \cite{Bullock:1996at}).
This $t_i$ should not be confused with $t_i$ of Sec. \ref{sec-form}.
The size of the region at $t_c$ is
\begin{equation}
\tilde {\bf R}(t_c) = {\bf R}_c = \delta_i^{-1/2} \tilde {\bf R}(t_i) = \delta_i^{-1/2} {\bf R}_i.
\end{equation}
The perturbed region will contract and, eventually, collapse, if it contains
enough matter to get over pressure forces, i.e., if its radius exceeds the Jeans length,
\begin{equation}
{\bf R}_c \gtrsim {\bf R}_{\rm Jeans} = c_s t_c \sim \frac{1}{\sqrt{3}} t_c.
\end{equation}
The corresponding requirement for the PBH formation is
\begin{equation}
\frac{1}{\sqrt{3}} t_c \lesssim {\bf R}_c \lesssim t_c
\label{tc}
\end{equation}
(the value of the upper bound is explained in [\cite{CarrHawking}, \cite{Carr:1975qj}]).
Dividing (\ref{tc}) by $t_c$, we obtain the expression for ${\bf R}_c/t_c$,
\begin{equation}
\frac{{\bf R}_c}{t_c} = \frac{{\bf R}_i}{t_i} \delta_i^{1/2}
\label{Rc}
\end{equation}
which scales, approximately, like a constant (independent on time) because
\begin{equation}
{\bf R}_i \sim \sqrt{t_i} \;\; , \;\; \delta_i \sim t_i.
\end{equation}
Evaluating (\ref{Rc}) at horizon crossing (where ${\bf R}_H=t_H$), we obtain from (\ref{tc})
the necessary condition for the collapse (and the PBH formation)
\begin{equation}
\frac{1}{3} < \delta_R^H < 1.
\end{equation}
Here we used the notation for the density contrast at horizon crossing introduced
in Sec. \ref{sec-form}.

The mass of a black hole formed in a process of the collapse is given by
\begin{equation}
M_{BH} = \frac{4 \pi}{3} {\bf R}_c^3 \rho(t_c),
\label{l0}
\end{equation}
where, from (\ref{Rc}) and (\ref{tcrho}),
\begin{equation}
{\bf R}_c = \frac{{\bf R}_i}{t_i} \delta_i^{1/2} t_c \;\; ; \;\; t_c = t_i \delta_i^{-1};
\label{l1}
\end{equation}
\begin{equation}
\rho(t_c) = \rho_i \left( \frac{t_i}{t_c} \right) ^ 2 = \rho_i \delta_i^2 .
\label{l2}
\end{equation}
Substituting (\ref{l1}) and (\ref{l2}) to (\ref{l0}), one obtains
\begin{equation}
M_{BH} = \frac{4\pi}{3} \left[ \frac{{\bf R}_i}{t_i} \delta_i^{1/2} \right]^3
t_i^3 \delta_i^{-1} \rho_i.
\label{ll}
\end{equation}

The whole expression (\ref{ll}), in its r.h.s., scales, approximately, like a constant
with time, similar with r.h.s. of Eq. (\ref{Rc}). Evaluating it at horizon crossing (${\bf R}_H=t_H$),
we obtain the connection between black hole mass, horizon mass $M_h$ and $\delta_R^H$, given in
Sec. \ref{sec-form},
\begin{equation}
M_{BH} = \frac{4 \pi}{3} (\delta_R^H)^{1/2} t_H^3 \rho_H = M_h (\delta_R^H)^{1/2}.
\end{equation}

Using the connection between $M_h$, $M$ and $M_i$ given in Eq. (\ref{MhMiM}) of Sec. \ref{sec-form},
one obtains the well-known formula for $M_{BH}$ (see, e.g., \cite{Kim:1996hr, BugaevD65, Kim:1999iv}):
\begin{eqnarray}
M_{BH} =  (\delta_R^H)^{1/2} M_i^{1/3} M^{2/3}, \;\;\;\;\;\;
\delta_R^H \approx \frac{1}{3} = \gamma. \;\;\;\;\;\;\;\;\;\;
\end{eqnarray}

\end{document}